\documentclass{elsart}

\usepackage{natbib}
\usepackage{graphics}
\usepackage{times}
\usepackage{epsfig}
\usepackage{amssymb}

\newcommand{\sauron}{{\texttt {SAURON}}}

\newcommand{\chandra}{{\texttt {Chandra}}}
\newcommand{\Oiii}{[{\sc O$\,$iii}]}

\newcommand{\Hb}{H$\beta$}
\newcommand{\Hi}{{\sc H$\,$i}}

\newcommand{\eg}{e.g.,}

\newcommand{\apj}{ApJ}
\newcommand{\apjl}{ApJ}
\newcommand{\apjs}{ApJS}
\newcommand{\aj}{AJ}
\newcommand{\mnras}{MNRAS}

\newcommand{\aap}{A\&A}

\def\spose#1{\hbox to 0pt{#1\hss}}
\def\lta{\mathrel{\spose{\lower 3pt\hbox{$\sim$}}
    \raise 2.0pt\hbox{$<$}}}
\def\gta{\mathrel{\spose{\lower 3pt\hbox{$\sim$}}
    \raise 2.0pt\hbox{$>$}}}

\newcommand{\placefigone}{
\begin{figure}
\begin{center}
\epsfig{file=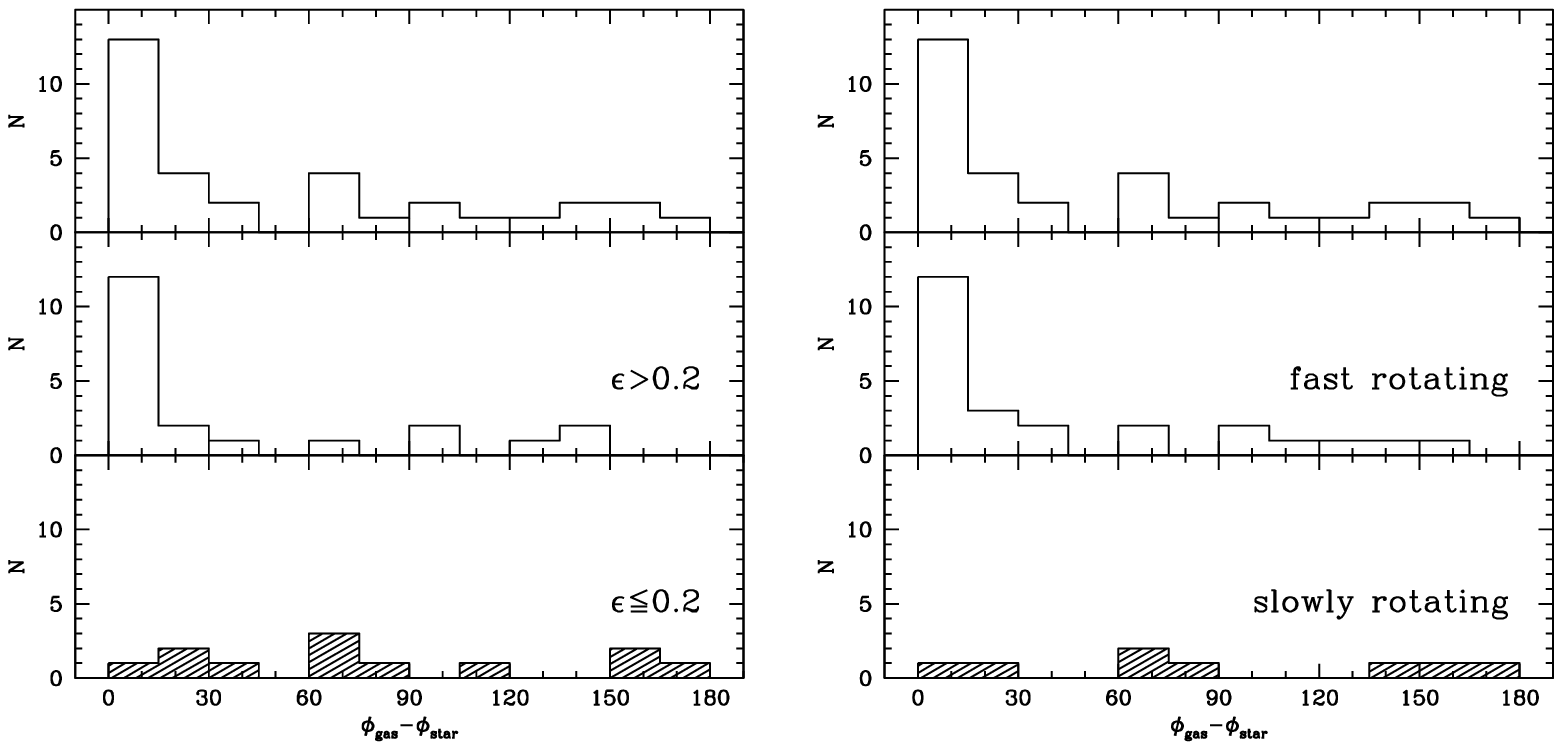, width=\textwidth, bb = 33 35 484 251}
\end{center}
\caption{
Distribution of the values for the kinematic misalignment between stars
and gas for \sauron\ early-type galaxies with clear stellar rotation
and well-defined gas kinematics. The distribution corresponding to all
the \sauron\ galaxies satisfying these conditions is shown in both top
panels, whereas the middle panels show the misalignments distribution
only for flatter and fast rotating objects (on the left and right,
respectively) and the lower panels only for rounder and slowly
rotating systems (on the left and right, respectively). Adapted from
\citet{Sar06}.}
\label{fig:Misal}
\end{figure}
}

\newcommand{\placefigtwo}{
\begin{figure}
\begin{center}
\epsfig{file=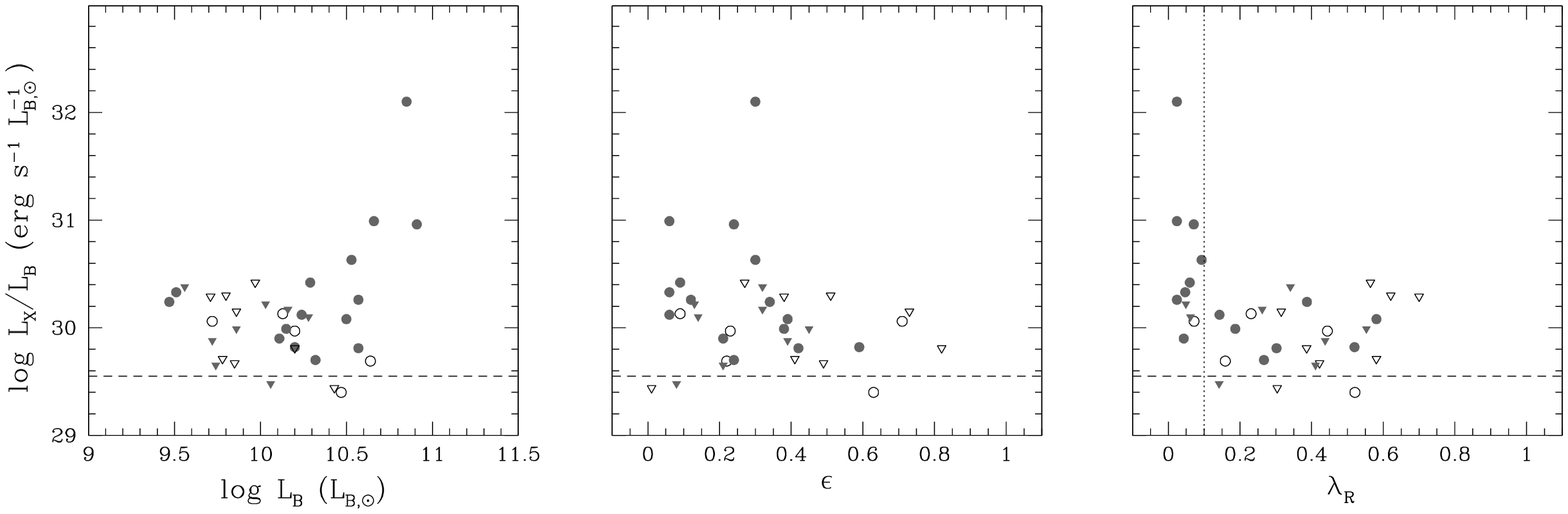, width=\textwidth, bb = 17 21 718 251}
\end{center}
\caption{X-ray properties of \sauron\ early-type galaxies.
From left to right the X-ray luminosity $L_X$, normalised to the total
blue-band luminosity $L_B$, is compared to $L_B$, the flattening
$\epsilon$ and the degree of rotational support as traced by the
$\lambda_R$ parameter of Emsellem et al. (2007). 
$L_X$ values (circles) or upper-limits (triangles) are {\tt ROSAT\/}
and {\tt Einstein\/} measurements from the compilation of o'Sullivan
et al. (2001), except for NGC524 and NGC5813, which were taken from
Pellegrini et al. (2005) and Bohringer et al. (2000),
respectively. These data cover three-quarters of the \sauron\
early-type sample. Filled and open symbols show elliptical and
lenticular galaxies, respectively. In each panel, the horizontal
dashed line shows the normalised X-ray luminosity expected from
stellar sources (o'Sullivan et al. 2001).
Notice the tight segregation of objects with high-$L_X/L_B$ to the
region with $\lambda_R < 0.1$ -- only slow-rotators can retain massive
X-ray halos.}
\label{fig:LbLx}
\end{figure}
}

\newcommand{\placefigthree}{
\begin{figure}
\begin{center}
\epsfig{file=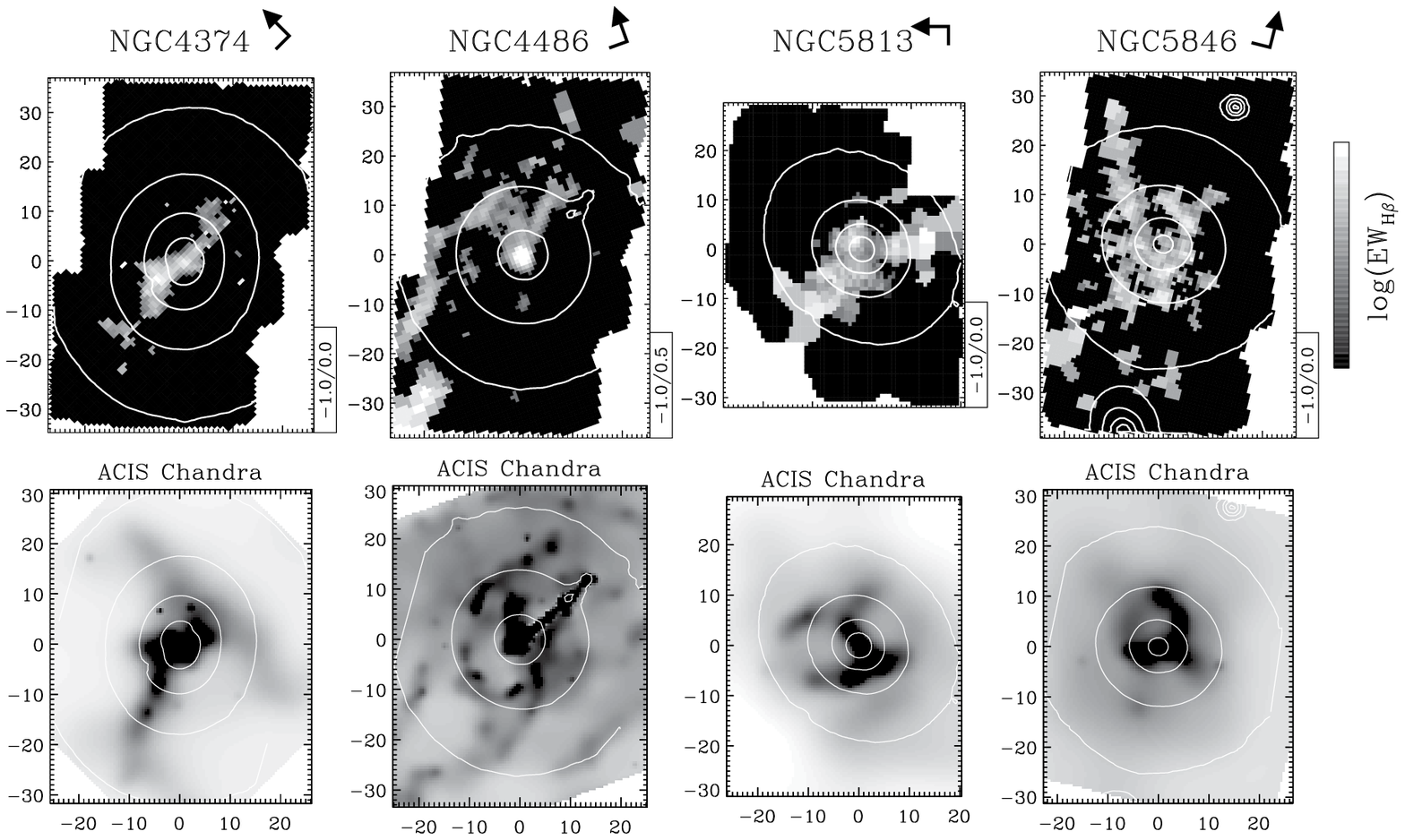, width=\textwidth}
\end{center}

\caption{Warm {\it versus\/} hot gas emission in four slowly-rotating \sauron\ galaxies. 
The top panels show the equivalent width of the \Hb\ line, in \AA\ and
on a logarithmic scale, whereas the lower panels show smoothed maps
for the X-ray emission observed with \chandra, in arbitrary
units. Except in only few regions, nebular emission is usually
associated to X-ray emitting features. For NGC~4486 and NGC~5846, this
connection was already discussed by \citet{Spa04} and \citet{Tri02},
respectively}
\label{fig:Xray}
\end{figure}
}

\begin{document}

\begin{frontmatter}



\title{On the origin and fate of ionised-gas in early-type galaxies: the \sauron\ perspective}


\author[UH]{Marc Sarzi}\thanks{E-mail: sarzi@star.herts.ac.uk},
\author[CRAL]{Roland Bacon},
\author[Oxford]{Michele Cappellari},
\author[Oxford]{Roger L. Davies},
\author[CRAL]{Eric Emsellem},
\author[ESTEC]{Jes\'us Falc\'on-Barroso},
\author[Oxford]{Davor Krajnovi\'c},
\author[ESA]{Harald Kuntschner},
\author[Sterrewacht]{Richard M. McDermid},
\author[Groningen]{Reynier F. Peletier},
\author[Sterrewacht]{Tim de Zeeuw},
\author[Princeton]{Glenn van de Ven}

\address[UH]{Centre for Astrophysics Research, University of Hertfordshire, United Kingdom}
\address[CRAL]{Universit\'e de Lyon 1, CRAL, Observatoire de Lyon, France}
\address[Oxford]{Sub-Department of Astrophysics, University of Oxford, United Kingdom}
\address[ESTEC]{European Space Research and Technology Center, The Netherlands}
\address[ESA]{ST-ECF, European Southern Observatory, Germany}
\address[Sterrewacht]{Sterrewacht Leiden, The Netherlands}
\address[Groningen]{Kapteyn Astronomical Institute, The Netherlands}
\address[Princeton]{Department of Astrophysical Sciences, Princeton, U.S.A.}

\begin{abstract}
By detecting ionised-gas emission in 75\% of the cases, the \sauron\
integral-field spectroscopic survey has further demonstrated that
early-type galaxies often display nebular emission. Furthermore, the
\sauron\ data have shown that such emission comes with an intriguing
variety of morphologies, kinematic behaviours and line ratios.
Perhaps most puzzling was the finding that round and slowly rotating
objects generally display uncorrelated stellar and gaseous angular
momenta, consistent with an external origin for the gas, whereas
flatter and fast rotating galaxies host preferentially co-rotating gas
and stars, suggesting internal production of gas.
Alternatively, a bias against the internal production of ionised gas
and against the acquisition of retrograde material may be present in
these two kinds of objects, respectively.
In light of the different content of hot gas in these systems, with
slowly rotating objects being the only systems capable of hosting
massive X-ray halos, we suggest that a varying importance of
evaporation of warm gas in the hot interstellar medium can contribute
to explain the difference in the relative behaviour of gas and stars
in these two kinds of objects.
Namely, whereas in X-ray bright and slowly rotating galaxies
stellar-loss material would quickly evaporate in the hot medium, in
X-ray faint and fast rotating objects such material would be allowed
to lose angular momentum and settle in a disk, which could also
obstruct the subsequent acquisition of retrograde gas.
Evidence for a connection between warm and hot gas phases, presumably
driven by heat conduction, is presented for four slowly rotating
galaxies with \chandra\ observations.
\end{abstract}


\end{frontmatter}

\section{Introduction}
\label{sec:intro}

Over the years, a number of imaging and spectroscopic studies have
contributed to end the preconception about early-type galaxies as
consisting of simple stellar systems \citep[see][for a
review]{Gou99}. Yet, if it is now accepted that elliptical and
lenticular galaxies often contain dust and display nebular emission, a
number of questions still remain unanswered.
What is the origin of the interstellar material in early-type galaxies
galaxies? Is it material lost by stars during their evolution or does
it have an external origin? And what is its fate? Does it cool down to
form stars or does it become hot, X-ray emitting gas?
Furthermore, what powers the observed nebular emission? Is it a
central active nucleus? Is the warm ($\sim\!10^4$K) gas ionised by the
hot ($\sim\!10^7$K) gas through heat conduction \citep[\eg][]{Spa93}?
Is the gas ionised by stars; young or old (\eg\ post-AGB)? Or is it
excited by shocks, as also proposed for low-ionisation nuclear
emission-line regions \citep{Dop95}.

The \sauron\ integral-field spectroscopic survey \citep{deZ02} has not
only further demonstrated that early-type galaxies often display
nebular emission, but also shown that such emission comes with a
variety of morphologies, kinematic behaviours and line ratios, which
suggests a rather complex picture for the origin, fate and ionisation
of the gas.
Using a new procedure that simultaneously fits both the stellar
spectrum and the emission lines, \citet[][hereafter Paper~V]{Sar06}
could indeed measure H$\beta$, \Oiii$\lambda\lambda$4959,5007, and
[{\sc N$\,$i}]$\lambda\lambda$5198,5200 emission lines down to
equivalent width values of 0.1\AA, uncovering extended emission in
75\% of the 48 elliptical and lenticular galaxies in the \sauron\
sample.
Across these objects, the gas emission is found in disks, in
filamentary structures, along lanes, or in rings and spiral arms.
The ionised-gas kinematics is rarely consistent with simple coplanar
circular motion, and generally display coherent motions with smooth
variation in angular momentum.
Finally, a considerable range of values for the \Oiii/\Hb\ ratio is
observed both across the sample and within single galaxies, which,
despite the limitations of this ratio as an emission-line diagnostic,
suggests that a variety of mechanisms is responsible for the gas
excitation in E and S0 galaxies.

This paper will focus primarily on the origin and fate of the ionised
gas in early-type galaxies. In \S~\ref{sec:gasstar} we will discuss
clues from the relative kinematic behaviour of gas and stars in
early-type galaxies, whereas in \S~\ref{sec:hotwarm} we will consider
some suggestive facts concerning the hot-gas content of the elliptical
and lenticular galaxies in the \sauron\ sample. Finally, we will draw
our conclusions in \S~\ref{sec:concl}.

\placefigone
\section{Clues from the kinematic decoupling between gas and stars}
\label{sec:gasstar}

The distribution of the values for the misalignment between the
angular momenta of gas and stars in early-type galaxies has often been
used to determine the relative importance of accretion events and the
internal production of gas through stellar mass-loss
\citep[e.g.,][]{Ber92}.

If the origin of the gas is external and gas is acquired from random
directions, an equal number of co- and counter-rotating gaseous and
stellar systems should be found, with a more or less pronounced
fraction of objects with gas in polar orbits depending on the intrinsic
shape of early-type galaxies \citep{Ste82}.
On the other hand, gas that is internally produced will rotate in the
same sense of its parent stars and only co-rotating gas and stars
should be observed.
In both cases, projection effects and the presence of triaxial systems
will lead to observe also intermediate values for the kinematic
misalignment, beside $0^\circ$, $90^\circ$ and $180^\circ$.
Overall, however, if the gas has an external origin the resulting
distribution for the kinematic misalignments should be {\it
symmetric\/} around $90^\circ$, with an equal number of counter- and
co-rotating gaseous and stellar systems, whereas if the origin of the
gas is internal the distribution of kinematic misalignments should be
{\it asymmetric\/}, with values mostly between $0^\circ$ and
$90^\circ$.

Using the early-type galaxies that in the \sauron\ sample display
clear stellar rotation and well-defined gas kinematics, in Paper~V we
have found a distribution for the gas-star kinematic misalignments
that is inconsistent with the prediction of either of these simple
scenarios.
As the top panels of Fig.~\ref{fig:Misal} show, half of the objects
display a kinematic decoupling that implies an external origin for the
gas, but the number of objects consistent with co-rotating gas and
stars exceeds by far the number of counter-rotating systems,
suggesting that internal production of gas has to be important.

The distribution of values for the kinematic misalignment between
stars and gas does not depend on Hubble type, galactic environment, or
galaxy luminosity \citep[Paper~V, but see also Fig.~2 of][]{Fal06a}.
It does, however, strongly depend on the apparent large-scale
flattening of galaxies.
Figure~\ref{fig:Misal} shows that the roundest objects in our sample
($\epsilon \le 0.2$) present a more symmetric distribution of
kinematic misalignments than flatter galaxies, which instead host
predominantly co-rotating stellar and gaseous systems.
Since for random orientations fairly round galaxies are likely to be
almost spherical and hence supported by dynamical pressure, rather
than by rotation, the degree of rotational support could also be
important to explain the observed dependency on galaxy flattening.
In \citet{Ems07} we assess the level of rotation support adopting a
quantity, $\lambda_{\rm R}$, that is closely related to the specific
angular momentum of a galaxy.
In the \sauron\ sample, galaxies with $\lambda_{\rm R} \le 0.1$ form a
distinct class characterised by little or no global rotation and the
presence of kinematically decoupled cores. Opposed to such slowly
rotating objects are galaxies that either display faster global
rotation or that are consistent with being systems supported by
rotation that are viewed at small inclinations.
Figure~\ref{fig:Misal} shows the distribution of kinematic
misalignments between gas and stars in fast and slowly rotating
galaxies according to the criterion of Emsellem et al. \citep[see
also][for an illustration of these two kinds of objects]{McD05}.
Consistent with our expectations, the two distributions are remarkably
different, as in the case of flat and round objects.

According to the previous first-order assumptions these results
suggest that external accretion of gaseous material is less important
than internal production of gas in flat and fast rotating galaxies,
whereas rounder and slowly rotating objects would acquire their gas
more often.
Such an interpretation is quite puzzling, however, for there is no
obvious reason why early-type galaxies would acquire or produce more
or less gas depending only on their apparent flattening or level of
rotation support.
The flat and fast rotating objects plotted in Fig.~\ref{fig:Misal} do
not live in a different environment than the round and slowly rotating
systems, and except for a few cases, all objects share old and evolved
stellar populations \citep{Kun06}.

This leads to consider alternative explanations for the observed
distribution of values for the kinematic misalignment between gas and
stars, starting from the assumption that all early-type galaxies can
both acquire and produce gaseous material. In this framework, what is
needed are mechanisms that would favour the accretion of prograde over
retrograde material in flatter and fast rotating galaxies, and remove
or hide the gas internally produced in rounder and slowly rotating
objects.

For instance, a pre-existing indigenous gaseous disk could obstruct
the acquisition of material with anti-parallel angular momentum. This
appears to be the case of late-type galaxies where indeed
counter-rotating gaseous and stellar systems are very seldom observed
(see Falc{\'o}n-Barroso et al. 2006b and Ganda et al. 2006, for a
\sauron\ view on early- and late-type spiral galaxies, respectively,
and Bertola \& Corsini 1999, for a review on the phenomenon of
counter-rotation).
On the other hand, the presence of a massive halo of hot, X-ray
emitting gas could render undetectable in the visible spectrum the gas
shed by stars as this evaporates in the hot gaseous medium
\citep[e.g.,][]{Mat90}.

\placefigtwo
\section{Clues from the X-ray properties}
\label{sec:hotwarm}

Following the previous considerations, it is interesting to note that
early-type galaxies display dramatically different behaviours when it
comes to their X-ray luminosity $L_X$.
Indeed, whereas in faint elliptical and lenticular galaxies X-ray
binary stars can account for the observed X-ray fluxes, the brightest
early-type galaxies tend to display truly extended and massive halos
of hot, X-ray emitting gas \citep[\eg][]{Fab92}. This can be
appreciated when $L_X$ is compared to the blue-band luminosity $L_B$,
as $L_X\sim L_B$ at low-luminosities and $L_X\sim L_B^2$ for $L_B \gta
3\times10^9 L_{B,\odot}$ \citep[\eg][]{OSu01}. The scatter in the
$L_X-L_B$ diagram is very large, however, and has been the subject of
many investigations \citep[see][for a review]{Mat03}. In particular,
Eskridge, Fabbiano \& Kim (1995a,b) found that S0s and flat Es show
lower X-ray luminosities than rounder elliptical galaxies of the same
optical luminosity, which prompted theoretical studies concerning the
r\^ole of intrinsic flattening or rotation in producing different
$L_X$ at a given $L_B$ \citep{Cio96,Bri96}. 

Fig.~\ref{fig:LbLx} shows, for the \sauron\ early-type galaxies\ with
global X-ray luminosity measurements, $L_X/L_B$ {\it versus \/} $L_B$,
$\epsilon$ and $\lambda_R$. Consistent with previous studies, also in
the \sauron\ sample the brightest X-ray halos are exclusively found
around the brightest and roundest objects, although there are also
galaxies with $\epsilon \lta 0.3$ and $L_B \gta 3\times10^{10}
L_{B,\odot}$ that exhibit normalised X-ray luminosities consistent with
stellar sources (dashed line). 
On the other hand, when $L_X/L_B$ is compared to the $\lambda_R$
parameter it is quite remarkable not only that high $L_X/L_B$ values
are found solely in galaxies with $\lambda_R < 0.1$, i.e. the
slow-rotators, but also that the previously mentioned round or bright
objects with faint normalised X-ray luminosities turned out to be fast
rotators.
In other words, $\lambda_R$ works better than the apparent flattening
or the total luminosity of a galaxy in separating objects with or
without extended X-ray halos.

The presence of massive halos of hot, X-ray emitting gas in rounder
and slowly rotating galaxies could explain why a conspicuous number of
co-rotating gaseous and stellar systems is not observed in this class
of objects, since the stellar-loss material would be efficiently
removed from the observed ionised-gas component of the interstellar
medium.

The conduction of heat from hot to warm medium can be observed when
the warm recipient is material that has been recently acquired and
that comes with sufficient column density.
Owing to X-ray images with high spatial resolution, evidence for such
interaction was presented by \citet{Tri02} and \citet{Spa04} as they
revealed a striking spatial coincidence between regions with
ionised-gas emission and specific features in the hot gas distribution
in the giant elliptical galaxies NGC~5846 and NGC~4486, respectively.
\citet{Spa04} also show that where hot and warm emission coincide the
temperature of the hot gas is lower than in the surrounding regions, as
expected in the presence of a heat sink.

Here we bring two additional examples of such interaction.
Fig.~\ref{fig:Xray} compares the distribution of the ionised-gas
emission with that of the hot, X-ray emitting gas for the four
slowly-rotating galaxies with the brightest X-ray halos in the
\sauron\ sample, which include NGC~5846 and NGC~4486.
Except in only few regions, we found that the \Hb\ emission is usually
associated to X-ray emitting features in the \chandra\ images.

If the fate of stellar-mass loss material in slowly-rotating galaxies
is to join the hot component of the interstellar medium, we suggest
that in the presence of a much less significant hot medium the stellar
ejecta may have the time to collide, loose angular momentum, and
settle on the galactic plane.
As such a gaseous disk could obstruct the subsequent acquisition of
retrograde gas material, the absence of bright X-ray halos in
fast-rotating galaxies would contribute to explain the excess
of objects with co-rotating gas and stars and the relative shortage of
counter-rotating gaseous and stellar systems.
Additionally, if the hot interstellar medium is heated primarily by the
dissipation of the kinetic energy of the stellar ejecta, it is likely
that part of the stellar angular momentum is transferred to the hot
gas. Although their existence remains controversial
\citep[][]{Han00,Die06}, rotating X-ray halos would exert larger
ram-pressure forces on material that is accreted with opposite angular
momentum to the stars and the hot gas, thus further obstructing the
accretion of retrograde material.

\placefigthree

\section{Conclusions}
\label{sec:concl}

Starting from our finding that 75\% of the \sauron\ early-type sample
galaxies show ionised-gas emission, and armed with the new tool of
Emsellem et al. to assess the level of rotation support in galaxies,
we have explored different venues for the origin and fate of the gas
in these systems.

In particular, we have focused on the puzzling finding that round and
slowly rotating objects generally display uncorrelated stellar and
gaseous angular momenta, whereas flatter and fast rotating galaxies
host preferentially co-rotating gas and stars.
At face value this result suggests that external accretion of
gaseous material is less important than internal production of gas in
flat and fast rotating galaxies, whereas rounder and slowly rotating
objects would acquire their gas more often.
There are however no obvious reasons to support this view, in
particular because these two kinds of galaxies inhabit similar
galactic environments.

Assuming that all early-type galaxies can produce and acquire gaseous
material, we have therefore considered the possibility that other
mechanisms could obstruct the acquisition of counter-rotating gas in
flat and fast rotating galaxies and remove or render undetectable the
stellar-loss material in round and slowly rotating objects.
Given the potential role of interaction between different phases
of the interstellar medium, and building on past findings that S0s and
flat Es show lower X-ray luminosities than rounder elliptical galaxies
of the same optical luminosity, we have compiled X-ray luminosity
measurements for our sample galaxies and found that only the objects
classified as slowly rotating on the basis of integral-field
observations can retain massive halos of hot gas.

In light of the different content of hot gas in slow and fast rotating
galaxies, we have suggested that the different importance that
evaporation of warm gas in the hot interstellar medium has in these
galaxies can contribute to explain the difference in the relative
behaviour of gas and stars in these two kinds of objects.
Namely, whereas in X-ray bright and slowly-rotating galaxies
stellar-loss material would quickly join the hot medium and escape
detection as ionised-gas, in X-ray faint and fast-rotating objects
such material would be allowed to lose angular momentum and settle on
a disk, which could also obstruct the further acquisition of
retrograde gas.
Thus, the evaporation of stellar ejecta in the hot medium and the
obstructing action of an indigenous disk would explain the relative
shortage of co-rotating and counter-rotating gaseous and stellar
systems in slow and fast rotating galaxies, respectively.

We note that the fast-rotating objects with the largest values for the
specific mass of ionised gas, that is, the ionised gas mass normalised
by the virial mass of the galaxy, tend to have a more uniform
distribution for the gas-star kinematic misalignments, consistent with
an external origin for their gas. Among the fast-rotating objects with
ionised-gas mass fractions above $2\times10^{-6}$, three have corotating
gaseous and stellar systems, one shows gas in polar orbits, and four
display retrograde gas motions.
A larger specific mass of ionised gas may reflect the need for
sufficiently large amounts of gas in order to overcome the obstructing
action of an indigenous disk, or the triggering of additional sources
of ionisation, such as shocks or star formation. As regards the
latter, McDermid et al. (this volume, their Fig.~2) indeed find
younger stellar population embedded in these objects.

A quantitative assessment of the relative importance of evaporation in
the hot interstellar medium and of dissipative processes, as well as
more constraints on the amount of cold gas at both small and large
scales (from CO and \Hi\ observations), will be needed to confirm the
picture proposed in this paper.

\ack

Marc Sarzi wishes to thank the organisers of the conference, in
particular Raffaella Morganti, for their support.





\begin{thebibliography}{}


\bibitem[Bertola et al.(1992)]{Ber92} Bertola, F., Buson, L.~M., \&
Zeilinger, W.~W.\ 1992, \apjl, 401, L79
\bibitem[Bertola \& Corsini(1999)]{Ber99} Bertola, F., \& Corsini,
E.~M.\ 1999, IAU Symp.~186: Galaxy Interactions at Low and High
Redshift, 186, 149
\bibitem[B{\"o}hringer et al.(2000)]{Bor00} B{\"o}hringer H., et al.,
2000, \apjs, 129, 435
\bibitem[Brighenti \& Mathews(1996)]{Bri96} Brighenti F., Mathews
W.~G., 1996, \apj, 470, 747
\bibitem[Ciotti \& Pellegrini(1996)]{Cio96} Ciotti L., Pellegrini S.,
1996, \mnras, 279, 240
\bibitem[de Zeeuw et al.(2002)]{deZ02} de Zeeuw, P.~T., et al.\ 2002,
\mnras, 329, 513
\bibitem[Diehl \& Statler(2006)]{Die06} Diehl, S., \& 
Statler, T.~S.\ 2006, \apj, submitted, astro-ph/0606215
\bibitem[Dopita \& Sutherland(1995)]{Dop95} Dopita, M.~A., \&
Sutherland, R.~S.\ 1995, \apj, 455, 468
\bibitem[Emsellem et al.(2007)]{Ems07} Emsellem, E., et al.\ 2007,
\mnras, submitted
\bibitem[Eskridge, Fabbiano, \& Kim(1995a)]{Esk95a} Eskridge P.~B.,
Fabbiano G., Kim D.-W., 1995a, \apjs, 97, 141
\bibitem[Eskridge, Fabbiano, \& Kim(1995b)]{Esk95b} Eskridge P.~B.,
Fabbiano G., Kim D.-W., 1995b, \apj, 442, 523
\bibitem[Fabbiano, Kim, \& Trinchieri(1992)]{Fab92} Fabbiano G., Kim
D.-W., Trinchieri G., 1992, \apjs, 80, 531
\bibitem[Falc{\'o}n-Barroso et al.(2006)]{Fal06a} Falc{\'o}n-Barroso,
J., et al.\ 2006a, New Astronomy Review, 49, 515
\bibitem[Falc{\'o}n-Barroso et al.(2006)]{Fal06b} Falc{\'o}n-Barroso,
J., et al.\ 2006b, \mnras, 369, 529
\bibitem[Ganda et al.(2006)]{Gan06} Ganda, K., 
Falc{\'o}n-Barroso, J., Peletier, R.~F., Cappellari, M., Emsellem, E., 
McDermid, R.~M., de Zeeuw, P.~T., \& Carollo, C.~M.\ 2006, \mnras, 367, 46 
\bibitem[Goudfrooij(1999)]{Gou99} Goudfrooij, P.\ 1999, ASP
Conf.~Ser.~163: Star Formation in Early Type Galaxies, 163, 55
\bibitem[Hanlan \& Bregman(2000)]{Han00} Hanlan, P.~C., \& Bregman,
J.~N.\ 2000, \apj, 530, 213 
\bibitem[Kuntschner et al.(2006)]{Kun06} Kuntschner, H., et 
al.\ 2006, \mnras, 369, 497
\bibitem[Mathews(1990)]{Mat90} Mathews, W.~G.\ 1990, \apj, 
354, 468 
\bibitem[Mathews \& Brighenti(2003)]{Mat03} Mathews W.~G., Brighenti
F., 2003, ARA\&A, 41, 191
\bibitem[McDermid et al.(2005)]{McD05} McDermid, R.~M., et al.\ 2006,
New Astr. Rev., New Astronomy Review, 49, 521
\bibitem[O'Sullivan, Forbes, \& Ponman(2001)]{OSu01} O'Sullivan E.,
Forbes D.~A., Ponman T.~J., 2001, \mnras, 328, 461
\bibitem[Pellegrini(2005)]{Pel05} Pellegrini S., 2005, \mnras, 364,
169
\bibitem[Sarzi et al.(2006)]{Sar06} Sarzi, M., et al.\ 2006, \mnras,
366, 1151
\aj, 121, 2928
\bibitem[Sparks et al.(1993)]{Spa93} Sparks, W.~B., Ford, H.~C., \&
Kinney, A.~L.\ 1993, \apj, 413, 531
\bibitem[Sparks et al.(2004)]{Spa04} Sparks, W.~B., Donahue, M.,
Jord{\' a}n, A., Ferrarese, L., \& C{\^ o}t{\' e}, P.\ 2004, \apj,
607, 294
\bibitem[Steiman-Cameron \& Durisen(1982)]{Ste82} 
Steiman-Cameron, T.~Y., \& Durisen, R.~H.\ 1982, \apjl, 263, L51 
\bibitem[Trinchieri \& Goudfrooij(2002)]{Tri02} Trinchieri, G., \&
Goudfrooij, P.\ 2002, \aap, 386, 472

\end{thebibliography}
\end{document}